\input harvmac
\input epsf
%
\def\journal#1&#2(#3){\unskip, \sl #1\ \bf #2 \rm(19#3) }
\def\andjournal#1&#2(#3){\sl #1~\bf #2 \rm (19#3) }

\def\frac#1#2{{#1\over#2}}

\def\inbar{\,\vrule height1.5ex width.4pt depth0pt}
\def\IC{\relax\hbox{$\inbar\kern-.3em{\rm C}$}}
\def\IR{\relax{\rm I\kern-.18em R}}
\def\IP{\relax{\rm I\kern-.18em P}}

%
%

\def\ap#1#2#3{Ann. Phys. {\bf #1} (#2) #3}

\catcode`\@=11
\def\slash#1{\mathord{\mathpalette\c@ncel{#1}}}
\overfullrule=0pt

\def\KK{{\cal K}}

\def\MM{{\cal M}}

\def\underrel#1\over#2{\mathrel{\mathop{\kern\z@#1}\limits_{#2}}}

\catcode`\@=12


%

\def\exp{{\rm exp}}


\def\[{[}
\def\]{]}

\def\comment#1{ }
\def\simlt{\mathrel{\lower2.5pt\vbox{\lineskip=0pt\baselineskip=0pt
             \hbox{$<$}\hbox{$\sim$}}}}
\def\simgt{\mathrel{\lower2.5pt\vbox{\lineskip=0pt\baselineskip=0pt
             \hbox{$>$}\hbox{$\sim$}}}}


\rightline{CERN-TH/2001-066, RI-07-00, Stanford ITP-01/04}
\Title{
\rightline{hep-th/0103033}
}
{\vbox{\centerline{Little String Theory at a TeV}}}
\centerline{{\it Ignatios Antoniadis}\foot{On leave of absence from
CPHT, Ecole Polytechnique, UMR du CNRS 7644.}${}^{,2}$,
{\it Savas Dimopoulos${}^{3}$ and Amit Giveon${}^{4}$}}
\bigskip
\smallskip
\centerline{${}^2$Theory Division, CERN}
\centerline{CH-1211, Geneva 23, Switzerland}
\smallskip
\centerline{${}^3$Physics Department, Stanford University}
\centerline{Stanford CA 94305, USA}
\smallskip
\centerline{${}^4$Racah Institute of Physics, The Hebrew University}
\centerline{Jerusalem 91904, Israel}
\bigskip

\noindent
We propose a framework where the string scale as well as
all compact dimensions are at the electroweak scale $\sim$
TeV$^{-1}$. The weakness of gravity is attributed to the
small value of the string coupling $g_s \sim 10^{-16}$,
presumably a remnant of the dilaton's runaway behavior,
suggesting the possibility of a common solution to the
hierarchy and dilaton-runaway problems. In spite of the
small $ g_s$, in type II string theories with gauge
interactions localized in the vicinity of NS5-branes, the
standard model gauge couplings are of order one and are
associated with the sizes of compact dimensions. At a TeV
these theories exhibit higher dimensional and stringy
behavior. The models are holographically dual to a higher
dimensional non-critical string theory and this can be used
to compute the experimentally accessible spectrum and
self-couplings of the little strings. In spite of the
stringy behavior, gravity remains weak and can be ignored
at collider energies. 
The Damour-Polyakov mechanism is an automatic consequence of our scenario 
and suggests the presence of a massless conformally-coupled scalar, 
leading to potentially observable deviations from Einstein's theory,
including violations of the equivalence principle.

\vfill
\Date{2/2001}
\newsec{Introduction}

\lref\scales{I. Antoniadis, Phys. Lett. {\bf B246} (1990) 377.}
\lref\w{E. Witten, hep-th/9602070, Nucl. Phys. {\bf B471} (1996) 135.}
\lref\l{J.D. Lykken, hep-th/9603133, Phys. Rev. {\bf D54} (1996) 3693.}
\lref\add{N. Arkani-Hamed, S. Dimopoulos and G. Dvali,
hep-ph/9803315, Phys. Lett. {\bf B429} (1998) 263; I.
Antoniadis, N. Arkani-Hamed, S. Dimopoulos and G. Dvali,
hep-ph/9804398, Phys. Lett. {\bf B436} (1998) 257. N.
Arkani-Hamed, S. Dimopoulos and G. Dvali, Phys.\ Rev.\
{\bf D59} (1999) 086004.}
\lref\ab{I. Antoniadis, C. Bachas, hep-th/9812093, Phys. Lett. {\bf
B450} (1999) 83.}
\lref\ggl{H. Georgi and S.L. Glashow, Phys. Rev. Lett. {\bf 327} (1974) 438.}%
\lref\dg{S. Dimopoulos and H. Georgi, Nucl. Phys. {\bf B192}
(1981) 150.}%
\lref\ap{I. Antoniadis and B. Pioline, hep-th/9902055,
Nucl. Phys. {\bf B550} (1999) 41; hep-ph/9906480.}
\lref\aha{For a review, see O. Aharony, hep-th/9911147,
and references therein.}%

An obstacle to building a unified theory of all forces is the
enormous disparity between the gravitational and other forces,
commonly referred to as the hierarchy problem. In the standard
framework of particle physics this is answered at the expense of
postulating an enormous energy desert separating the gravitational
from the electroweak scale \ggl. The supersymmetric version of this
picture  \dg , called the supersymmetric standard model, has had a
quantitative success: the unification prediction of the value of
the weak mixing angle \dg, subsequently confirmed by the LEP and
SLC experiments.  This makes it tempting to believe in the
unification of the non-gravitational forces at a large energy
scale $\sim 10^{16}$ GeV. Nevertheless,  this picture leaves many
fundamental questions unanswered. There are
125 parameters in the supersymmetric standard model that remain
unexplained. These include the masses of the three generations of
particles and, above all, the incredible smallness of the
cosmological constant. This suggests that there are enormous gaps
in our understanding of Nature at low energies and that perhaps we
will need a radical revision of our fundamental view of the world
at low energies, at least with respect to gravity.

On the other hand, string theory provides the only known framework for
quantizing gravity. The cost is to replace our fundamental concept of
point particles by extended objects whose quantum consistency requires
the existence of extra dimensions. One of the important consequences of
the recent theoretical progress on the non-perturbative dynamics of
string theories is that the string and compactification scales are not
necessarily tied to the four-dimensional Planck mass \refs{\scales,\w,\l}.
This opens the exciting possibility that string physics may become relevant
at much lower energies with spectacular new effects in future accelerators.

Such a possibility can also be used to explain the hierarchy problem,
motivated by the following string theoretic expression for the
four-dimensional (4d) Planck mass \add:
\eqn\mp{M_P^2={1\over g_s^2}M_s^8 V_6~,}
where $g_s$ is the string coupling, $M_s$ the string
scale and $V_6$ the volume of the six-dimensional internal space.
This relation shows that it is possible that there is only one
fundamental scale in the universe, the electroweak scale $\sim$
TeV, where all forces of nature, including gravity, unify and
therefore $M_s \sim$ TeV. Then the enormity of the Planck scale
can be accounted for in two distinct ways:\hfill\break

(1) A non-stringy way \add:\hfill\break This is realized if $V_6$ is
enormously larger than the fundamental scale while keeping $g_s$
of order unity. In order to make such large dimensions consistent with
observations, gauge interactions should be localized on branes transverse
to them. A natural framework for realizing this scenario is type I string
theory with the Standard Model (SM) confined on a collection of
D$p$-branes. Perturbative calculability requires the $p-3$
longitudinal dimensions to be compactified near the (TeV) string
scale, while the $9-p$ transverse dimensions should be much larger in
order to account for the observed weakness of gravitational interactions.

While this scenario can be naturally imbedded in
type I string theories, it does not require string theory for its
implementation at low energies, below the (TeV) scale of quantum gravity.
The physical mechanism is the dilution of the strength of gravity by
spreading it into extra dimensions, which could have been invented by
Gauss two centuries ago. The hierarchy problem now turns into one of
explaining dynamically the large magnitude of $V_6$. \hfill\break

(2) A stringy way:\hfill\break This is realized by taking $V_6$ to
be of the order of the fundamental scale $\sim{\rm TeV}^{-6}$ and
attributing the enormity of the Planck mass to a tiny 
$g_s \sim 10^{-16}$ \ap\ (see also \l).~\foot{Actually,
the value of $g_s$ is more likely $10^{-14}$ \ap,
corresponding in eq. \mp\ to a volume $V_6\simeq (2\pi)^6$ of a
toroidal compactification with all radii fixed at the string length.} 
The hierarchy problem is now equivalent to
understanding the smallness of $g_s$, or equivalently the large
value of the dilaton field in our universe.
\hfill\break

Starting with \add, possibility (1) has been explored extensively
in the last three years. Our objective in this paper is to study
the second logical possibility (2) which is stringy in nature, at
least in the sense that it involves $g_s$,  and gives a new
perspective to the hierarchy and other problems in physics. A
fundamental question now becomes whether a string theory with such
a small $g_s$ can contain the ordinary gauge interactions whose
dimensionless couplings are of order unity. Fortunately the answer
is yes in the context of special type II
string theories whose gauge interactions are localized in the vicinity of
NS5-branes, which we will utilize here. In these theories gauge couplings
are given by the geometric sizes of new dimensions and are
non-vanishing even if $g_s$ vanishes.

\lref\rs{L. Randall and R. Sundrum, hep-ph/9905221,
Phys. Rev. Lett. {\bf 83} (1999) 3370.}
\lref\mal{J.M. Maldacena, hep-th/9711200,
Adv. Theor. Math. Phys. {\bf 2} (1998) 231.}
\lref\dap{T. Damour and A.M. Polyakov, Nucl. Phys. {\bf B423} (1994)
532; T.~Damour and A.~Vilenkin, Phys.\ Rev.\ {\bf D53}
(1996) 2981, hep-th/9503149.}%
\lref\seiberg{N. Seiberg, hep-th/9705221,
Phys. Lett. {\bf B408} (1997) 98.}%
\lref\brs{M.~Berkooz, M.~Rozali and N.~Seiberg,
Phys.\ Lett.\ {\bf B408} (1997) 105, hep-th/9704089.}%
\lref\dvv{R. Dijkgraaf, E. Verlinde and H. Verlinde,
Nucl. Phys. {\bf B486} (1997) 77, hep-th/9603126;
Nucl. Phys. {\bf B486} (1997) 89, hep-th/9604055;
Nucl. Phys. {\bf B506} (1997) 121, hep-th/9704018.}%
\lref\lms{A.~Losev, G.~Moore and S.L.~Shatashvili,
Nucl.\ Phys.\ {\bf B522} (1998) 105, hep-th/9707250.}%

In the limit of vanishing $g_s$ \seiberg\ one obtains a theory
without gravity, the so-called Little String Theory (LST);
it was introduced in \refs{\brs,\seiberg}
(see also \dvv,\lms, and for a review see \aha\ and references therein).
LST is  a partial string theory; although it
does not include gravity, it has string excitations and
therefore is not a normal local field theory. It is an
intermediate logical possibility between full-fledged
string theory and field theory. The main objective of our
paper is to point out that this intermediate possibility
can be realized at the experimentally accessible energy of
$\sim$ TeV and give us an alternate way to address the
hierarchy problem which connects it with the
dilaton-runaway problem. This therefore interpolates
between the TeV-strings framework \add, which has full
string theory at a TeV, and other field-theoretic
possibilities for TeV physics such as supersymmetry,
technicolor or warped compactifications \rs.

We propose three closely related frameworks for building realistic
theories with little strings at a TeV. Their common feature
is the existence of closed little strings with $\sim$ TeV
tension, whose self interaction and spectroscopy can be
computed in some cases. In addition there can be string
excitations of ordinary particles, with either the same or
different tension, as well as KK and winding modes
associated with $\sim{\rm TeV}^{-1}$ size dimensions.

An unexpected bonus of the framework is that the
Damour-Polyakov mechanism, based on the universality of the
dilaton coupling functions and normally considered
improbable, becomes automatic. It may lead to small but
potentially observable deviations from the equivalence
principle \dap.

In section 2, we discuss mass scales and couplings in type
II string theories, and define our general framework. In
section 3, we recall the possible descriptions of little
string theories and in particular the double scaling limit
which defines a sensible perturbation theory. In section 4,
we touch on some basic phenomenological consequences of the
framework. Section 5 addresses the hierarchy problem and
suggests ways in which the dilaton field can have a
naturally large value in our universe. In section 6, we
remark on a possible implication of our framework for the
cosmological constant problem and other topics.

\newsec{Mass Scales and Couplings}

In every perturbative string theory, gravity arises from closed strings
that propagate in ten dimensions. As a result, the 4d Planck mass is
given by eq. \mp. Here, all internal dimensions are taken to
be larger than the string length $l_s\equiv M_s^{-1}$ by a suitable
choice of T-dualities; thus, in this convention, all closed string
winding modes are heavier than the string scale. The strength of gravity
at energies above all compactification scales and below the string scale,
$V_6^{-1/6}<E<M_s$, is determined by the ten-dimensional Planck mass
\eqn\mstar{M_{10}^8={M_P^2\over V_6}={M_s^8\over g_s^2}~,}
which can be obtained by summing over all KK graviton
excitations yielding a suppression proportional to
$(E/M_{10})^8=(E/M_s)^8g_s^2$. It follows that at energies
of order the string scale gravitational interactions are
controlled by the string coupling $g_s$.

\lref\geen{For a review, see W. Lerche, hep-th/9611190,
Nucl. Phys. {\bf B55} (1997) 83;
A. Klemm, hep-th/9705131; P. Mayr, hep-th/9807096,
Fortsch. Phys. {\bf 47} (1999) 39.}%

On the other hand, in type II theories non-abelian gauge interactions
arise non-perturbatively localized on (Neveu-Schwarz) NS5-branes,
corresponding in the simplest case to D-branes stretched between the
NS5-branes. In a T-dual picture, non-abelian fields in (supersymmetric) type
IIA (IIB) theories emerge from D2 (D3) branes wrapping around collapsing
2-cycles of the compactification manifold \geen. Such 2-cycles are localized
in a subspace of dimension 4 that defines (upon T-duality) the transverse
position of the NS5-branes where gauge interactions are confined.
Furthermore, the four-dimensional Yang-Mills (YM) coupling is determined
by the geometry of the two-dimensional compact space along the
NS5-branes, independently of the value of the string coupling $g_s$.

{}For instance, let us consider a stack of NS5-branes extended in the
directions
$X^{0,1,2,3,4,5}$, where $X^{0,1,2,3}$ define our 1+3 dimensional
spacetime. The extra two longitudinal directions $X^{4,5}$
are compactified on a rectangular torus $T^2$ with radii $R_{4,5}$,
while the four transverse directions $X^{6,7,8,9}$
are compactified on a manifold with size $R_t$.
The four-dimensional gauge coupling is then given by
\eqn\gymf{\eqalign{{\rm type \,\, IIA:}&\qquad\qquad\qquad
g_{YM}^2={R_4\over R_5}~,\cr
{\rm type \,\, IIB:}&\qquad\qquad\qquad g_{YM}^2={l_s^2\over R_4R_5}~.\cr}}

In summary, in type II theories gravitational interactions
are controlled by the string coupling, while gauge
interactions are governed by geometrical moduli along the
5-branes where they are confined. This is in contrast with
type I theories, where the string coupling determines also
the strength of gauge interactions confined on D-branes and
is therefore fixed to be of order one~\foot{Here we drop
factors of $\pi$ and for numerical estimates we use
$g_{YM}\simeq 0.1$.}; thus, in type I theories
gravitational interactions become strong at the string
scale.

It follows that the type
II string scale can be lowered at the TeV scale without
introducing extra large transverse dimensions, but instead
a tiny string coupling to account for the hierarchy
$M_s/M_P$ \ap. In this case, the physics around the string
scale is described approximately by a theory without
gravity obtained in the weak coupling limit $g_s\to 0$ \seiberg.
This theory, which is defined in the limit of coincident
NS5-branes with vanishing string coupling, is called little
string theory (LST) \refs{\seiberg - \aha}.

This theory lives in six dimensions and
contains two sectors. The charged (non-abelian) sector
confined on the NS5-branes and a neutral sector of closed
fundamental strings trapped in the vicinity of the
NS5-branes. In section 3, we review the main properties of
these theories, while in section 4 we discuss their
phenomenological consequences when the fundamental string
scale is in the TeV region.

One may ask the question whether a tiny string coupling can
be described alternatively, via some duality,
in terms of large dimensions in the context of M-theory.
Indeed, it was shown that the
weakly coupled type II string compactified for instance on
$K3\times T^2$, with all compactified dimensions of string
size, provides a dual description to the strongly coupled
heterotic theory compactified on $T^4\times T^2$ with the
four dimensions of $T^4$ having the heterotic string size
$l_H\sim M_P^{-1}$ while the two dimensions of $T^2$ being
much larger, of the order of the type II string length \ap.
If the type II string scale $M_s\sim$ TeV, its string
coupling $g_s\sim M_s/M_P\simeq 10^{-16}$, while the
heterotic coupling is huge $g_H\sim l_s/l_H\sim 1/g_s$.
Using heterotic -- M-theory duality, one can find an
alternative description in terms of M-theory compactified
on the eleventh-dimensional interval $S^1/Z_2\times
T^4\times T^2$. The M-theory length scale
$l_M=g_H^{1/3}l_H$, so that $l_M^{-1}\sim 10^{14}$ GeV and
the size of the eleventh dimension $R_{11}=g_H l_H\sim{\rm
TeV}^{-1}$. Thus, there are three large dimensions at the
TeV ($R_{11}$ and $T^2$) and four small dimensions of
Planck length (corresponding to $T^4$) which invalidate the
effective field theory description and makes this M-theory
interpretation of no practical use.

\newsec{LST and D-branes in LST}

In this section we recall some aspects of LST and its
weakly coupled version (DSLST).
Using the idea of holography we review a dual description
of the theory as a ``non-critical'' string,
which allows to compute the spectrum
and couplings in some appropriate regime.
Moreover, this description provides a geometric
set up which has some analogy with the scenario of \rs.
We shall emphasize the similarities in this section, and we will make
some remarks at the end of the paper.
Finally, we also consider a different family of
theories, on D-branes in LST, which shares some
of the properties of LST.

\subsec{Little String Theory}

\lref\abks{O. Aharony, M. Berkooz, D. Kutasov and N. Seiberg,
hep-th/9808149, JHEP {\bf 9810} (1998) 004.}%

{}For simplicity, we first consider the six-dimensional LST
(for a review, see \aha\ and references therein).
One way to define this theory is the following.
We start with a stack of $k$ NS fivebranes in type II string theory
with a string coupling $g_s$, and take the
limit $g_s\to 0$. In this limit bulk degrees of freedom, including gravity,
are decoupled, and one is left with a six-dimensional theory
of strings without gravity.
An alternative description is to consider the $g_s\to 0$
limit of type II on a singular K3 manifold.
The two definitions are related by a T-duality.

This LST has the following properties \aha\ (in the conventions
of the first definition):
\item{(i)}
It has a unique scale $M_s$ -- the string mass scale of the
original type II string theory.
\item{(ii)}
In type IIA, the low energy theory is an $N=(2,0)$ six-dimensional SCFT,
while in type IIB it is an $N=(1,1)$, $SU(k)$ gauge theory with
a gauge coupling $g_{YM}\sim 1/M_s$.
\item{(iii)}
It has a Hagedorn density of states and the Hagedorn
temperature is $T_H= M_s/(2\pi\sqrt{k})$.
\item{(iv)}
It is argued \abks\ that this theory is ``holographically'' dual to
a higher dimensional string theory (with gravity):
the type II string on
\eqn\mmm{\MM=R^{5,1}\times R_{\phi}\times SU(2)_k~.}
Here $R^{5,1}$ is the $5+1$ dimensional Minkowski space in
the directions of the worldvolume of the fivebranes.
$R_{\phi}$ is the real line parameterized by a scalar
$\phi$, with a linear dilaton
\eqn\lidi{\Phi=-\sqrt{1\over 2k}\phi~.}
$\phi$ is related to the radial direction $r$ of the $R^4$
space transverse to the fivebranes: $\phi\sim \log r$.
The $SU(2)_k$ is a level $k$ WZW SCFT on the $SU(2)\simeq S^3$ background,
where this three sphere is related to the angular coordinates of
the transverse $R^4$.
This background is obtained in the near horizon limit of the fivebranes,
and describes the SCFT on the infinite ``throat'' (see figure 1(a)).
The fivebranes might be thought of as sitting deep down the throat,
namely, at $\phi\to -\infty$ ($r\to 0$) where the theory is strongly
coupled ($\exp(2\Phi)$ is large \lidi).
On the other hand, as $\phi\to\infty$ ($r\to\infty$, towards the
decoupled asymptotically flat space far from the fivebranes)
the theory is weakly coupled.
\item{(v)}
Off-shell observables in LST correspond to on-shell observables
in string theory on $\MM$ \mmm.
Observables in the theory correspond to non-normalizable vertex
operators, namely,
those whose wave function is exponentially supported at the weak coupling
regime $\phi\to\infty$.
There are also $\delta$ function normalizable operators whose role
in the theory is less clear.
The latter form a continuum of states, whose contribution to
the density of states is a small fraction.

\lref\gk{A. Giveon and D. Kutasov, hep-th/9909110,
JHEP {\bf 9910} (1999) 034.}%
\lref\gktwo{A. Giveon and D. Kutasov, hep-th/9911039,
JHEP {\bf 0001} (2000) 023.}%
\lref\efrw{S. Elitzur, A. Forge and E. Rabinovici,
Nucl. Phys. {\bf B359} (1991) 581;
G. Mandal, A.M. Sengupta and S.R. Wadia,
Mod. Phys. Lett. {\bf A6} (1991) 1685;
E. Witten, Phys. Rev. {\bf D44} (1991) 314.}%
\lref\sfetsos{K. Sfetsos, hep-th/9811167, JHEP {\bf 0001} (1999) 015;
hep-th/9903201.}%

\noindent
The holographic description above is useful to identify
observables and their properties under the symmetries of the theory.
However, correlation functions cannot be computed in perturbation
theory because they are sensitive to the strong coupling regime
down the throat.
To resolve this strong coupling problem we shall ``chop''
the strong coupling regime of the throat (see figure 1).

\vskip 1cm
\centerline{\epsfxsize=100mm\epsfbox{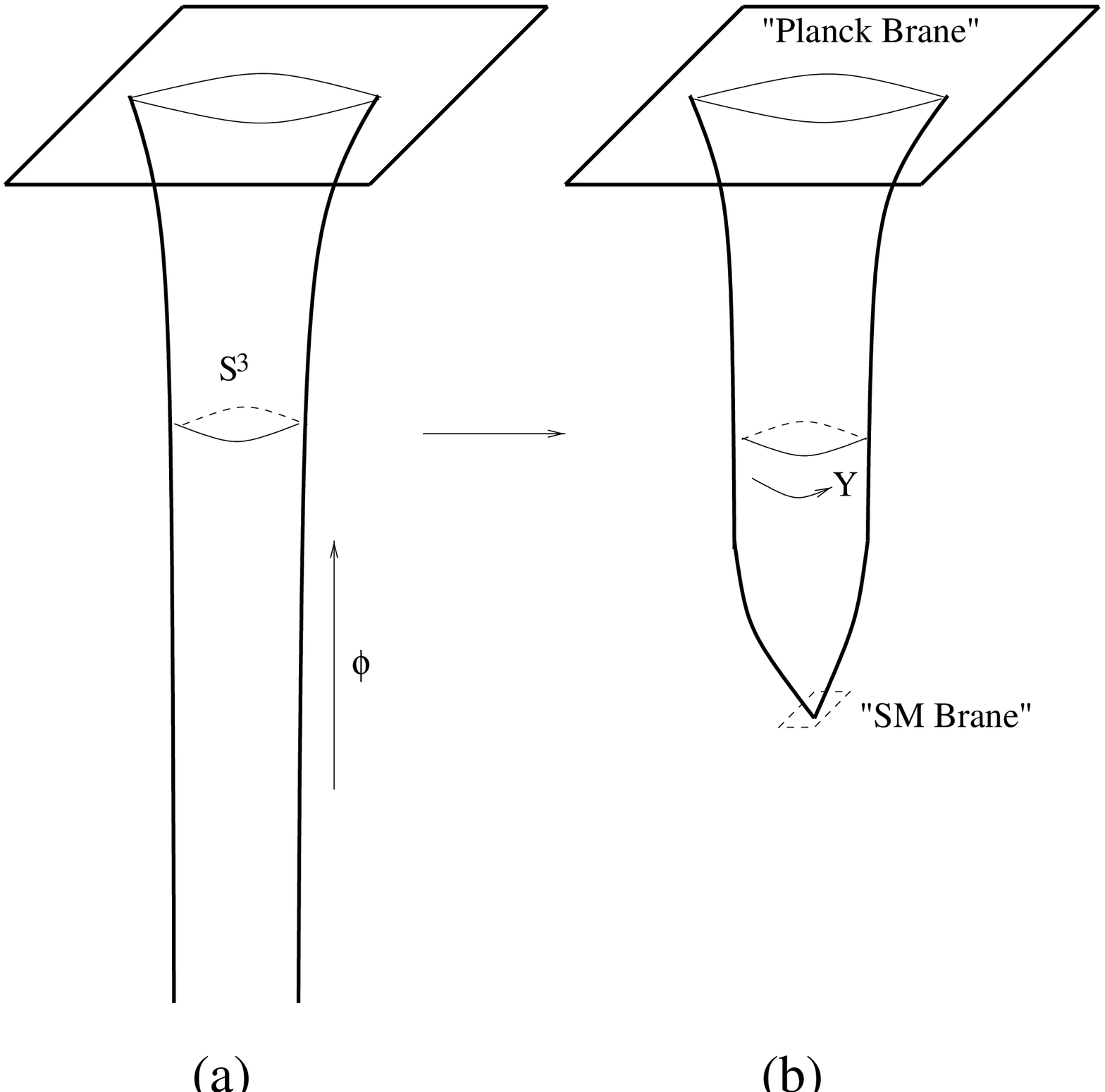}}
\vskip .1in
\noindent
Fig.\ 1: {\it (a) The infinite throat background dual to strongly
coupled LST; (b) The strong coupling region is chopped into a cigar-like
geometry, whose tip is associated with a SM brane while the asymptotic
region is associated with a Planck brane.}
\vskip .5cm

It is convenient to decompose the $SU(2)_k$ SCFT on $S^1_Y
\times SU(2)_k/U(1)$, where $S^1_Y$ is the Cartan
sub-algebra of $SU(2)$, parameterized by a scalar $Y$, and
the $SU(2)_k/U(1)$ quotient SCFT is equivalent to a level
$k$, $N=2$ minimal model. Then the throat SCFT becomes the
product of an infinite cylinder $R_\phi \times S^1_Y$ (with
a linear dilaton) times an $N=2$ minimal model:
\eqn\thr{R_\phi\times S^3
\simeq R_\phi \times S^1_Y \times SU(2)_k/U(1)~.}
One way \gk\ to chop the strong coupling regime of the throat
is to replace the infinite cylinder $R_\phi \times S^1_Y$
with the semi-infinite cigar \efrw\ SCFT $SL(2)_k/U(1)$ (see figure 1(b)):
\eqn\cigar{\MM=R^{5,1}\times R_{\phi}\times S^1_Y \times
{SU(2)_k\over U(1)}\to
R^{5,1}\times {SL(2)_k\over U(1)}\times {SU(2)_k\over U(1)}~.}
This corresponds to separating the $k$ NS fivebranes on a transverse
circle of radius $L$ in the double scaling limit $g_s,L\rightarrow 0$
such that $g_s/L$ is held fixed \refs{\sfetsos,\gk}.
The string coupling takes its maximal
value at the tip of the cigar where
\eqn\gtip{g_s(tip)\equiv g_{lst}\sim {g_s\over LM_s}~,}
while it approaches $0$ as one goes away from the tip
($\phi\to\infty$) along the radial direction $\phi$ of the
cigar. The scalar $Y$ parameterizes the angular direction
of the cigar whose radius is $R_{cigar}\sim \sqrt{k}/M_s$
asymptotically. The separation of the fivebranes introduces
another scale in the theory. In type IIB it is the mass of
a gauge boson corresponding to a D-string stretched between
two NS5-branes, giving rise to a charged particle in the
low energy $SU(k)$ gauge theory with mass
\eqn\mwb{M_W^{IIB}\sim T_{D1}L={M_s^2L\over g_s}={M_s\over g_{lst}}~,}
where $T_{D1}=M_s^2/g_s$ is the D-string tension.
One may regard the above as chopping the infinite throat
by SM branes (separated on a circle) near the tip of the cigar
(see figure 1(b)).

\lref\ver{H. Verlinde, hep-th/9906182,
Nucl. Phys. {\bf B580} (2000) 264.}%

So far we have considered the theory decoupled from gravity.
The decoupling limit corresponds, in particular, to the limit
$M_P\sim M_s/g_s\to\infty$.
To keep gravity at the finite (although large) scale $M_P$
observed in nature, we should relax the limit $g_s\to 0$
although the string coupling is still very small, as discussed in section 2.
One may regard this as chopping the weak coupling regime of
the semi-infinite cigar -- the other side of the original throat --
by a Planck brane~\foot{More precisely, it is done by keeping the
asymptotically flat regime ``glued'' to the throat, with the
four-dimensional space transverse to the NS5-branes being compactified
on $T^4/Z_2$ (similar to \ver) and with the appropriate number of
NS5-branes as required by global issues.} (see figure 1(b)).
We shall work in a scenario where $M_s, M_W^{IIB}\ll M_P$ and, therefore,
the effects of gravity for $E\sim M_s$ are negligible.

\lref\mo{N. Seiberg and E. Witten, hep-th/9903224,
JHEP {\bf 9904} (1999) 017; J. Maldacena and H. Ooguri, hep-th/0001053.}%
\lref\ags{R. Argurio, A. Giveon and A. Shomer, hep-th/0009242,
and references therein.}%

This Double Scaled LST (DSLST) \gk\
has a weak coupling expansion parameter
$g_{lst}=M_s/M_W^{IIB}$ when $M_W^{IIB}>M_s$
(we may however keep $M_W^{IIB}\ll M_P$).
This allows one, in principle, to compute correlation functions
perturbatively for processes at energies even larger than $M_s$
(as long as they are sufficiently lower than $M_W^{IIB}$).
On-shell correlators in the string theory
\cigar\ correspond, via holography, to off-shell Green's functions
in the six-dimensional spacetime theory. {}From the
analytic structure of the two point functions one can read
the physical spectrum while the three point functions give
rise to the couplings of physical states, via the LSZ
reduction. The two and three point functions were computed
in \refs{\gk,\gktwo}, with the following results:
\item{1.}{\it 2-p-f:}
The two point functions have a series of single poles, from
which one can read the mass spectrum, followed by a branch
cut (the poles correspond to the principal discrete series
in the unitarity range of the $SL(2)/U(1)$ SCFT while the
branch cut is due to the principal continuous series). The
massless states correspond to photon multiplets in the low
energy theory. They are followed by a discrete spectrum
organized into Regge trajectories due to string
excitations. The interpretation of the continuum is less
clear, and is probably associated with ``long strings''
(see \refs{\mo,\ags} and references therein). When $g_s$ is
finite the continuum in the spectrum is discretized.
\item{2.}{\it 3-p-f:}
The three point couplings allow, in particular, the decay
of a massive discrete state into two massless states.
Hence, one expects the stringy states to affect the form factor
of the ``photon'' at energies of the order $M_s$.

\lref\gkp{A. Giveon, D. Kutasov and O. Pelc, hep-th/9907178,
JHEP {\bf 9910} (1999) 035.}%
\lref\pelc{O. Pelc, hep-th/0001054, JHEP {\bf 0003} (2000) 012.}%
\lref\grka{M. Gremm and A. Kapustin, hep-th/9907210,
JHEP {\bf 9911} (1999) 018;
K. Benakli and Y. Oz, hep-th/9910090, Phys. Lett. {\bf B472} (2000) 83.}%

\noindent
Four-dimensional theories (at low energy)
can be constructed in various ways, for instance:
\item{(i)}
By compactifying two directions longitudinal to the fivebranes on
a two torus, as described in section 2.
The theory at energies below $M_s$ and the compactification scale
is an $N=4$, $SU(k)$, four-dimensional gauge theory.
\item{(ii)}
Four-dimensional LST whose low energy limit is an $N=2$
SCFT in the moduli space of pure $N=2$, $SU(n)$ gauge theory
can be studied by considering the near horizon of a fivebrane
wrapping a (singular) Riemann surface. Its holographic dual
is \gkp\ a type II string on
\eqn\four{\MM=R^{3,1}\times {SL(2)_k\over U(1)} \times {SU(2)_n\over U(1)}~,}
with $k={2n\over n+2}$.
\item{(iii)}
Richer four-dimensional LSTs can be obtained by replacing the
level $n$, $N=2$ minimal model in \four\ with a richer
$N=2$ SCFT. For instance, replacing ${SU(2)_n\over U(1)}$
by $[{SU(2)_n\over U(1)}\times {SU(2)_n\over U(1)}]/Z_n$,
with $k=n/2$,
leads \pelc\ to an $N=2$ SCFT with quark flavors.
\item{(iv)}
Theories with $N=1$ supersymmetry can be obtained by variations
of the theories above, for instance, by orbifolding and/or by
considering the decoupled theory on fivebranes
in the heterotic string \grka.

\subsec{Theories on D-branes in LST}

\lref\gkrev{A. Giveon and D. Kutasov, hep-th/9802067,
Rev. Mod. Phys. {\bf 71} (1999) 983.}%
\lref\egk{S.~Elitzur, A.~Giveon and D.~Kutasov,
Phys.\ Lett.\ {\bf B400} (1997) 269, hep-th/9702014.}%

In this subsection we discuss the 3+1 dimensional theory
on D4-branes stretched between NS5-branes when the string
scale $M_s$ is set, say, around 1 TeV.
In particular, in such theories we will be able to discuss
the spectrum and couplings of charged particles in theories
with TeV strings without gravity.

\vskip 1cm
\centerline{\epsfxsize=100mm\epsfbox{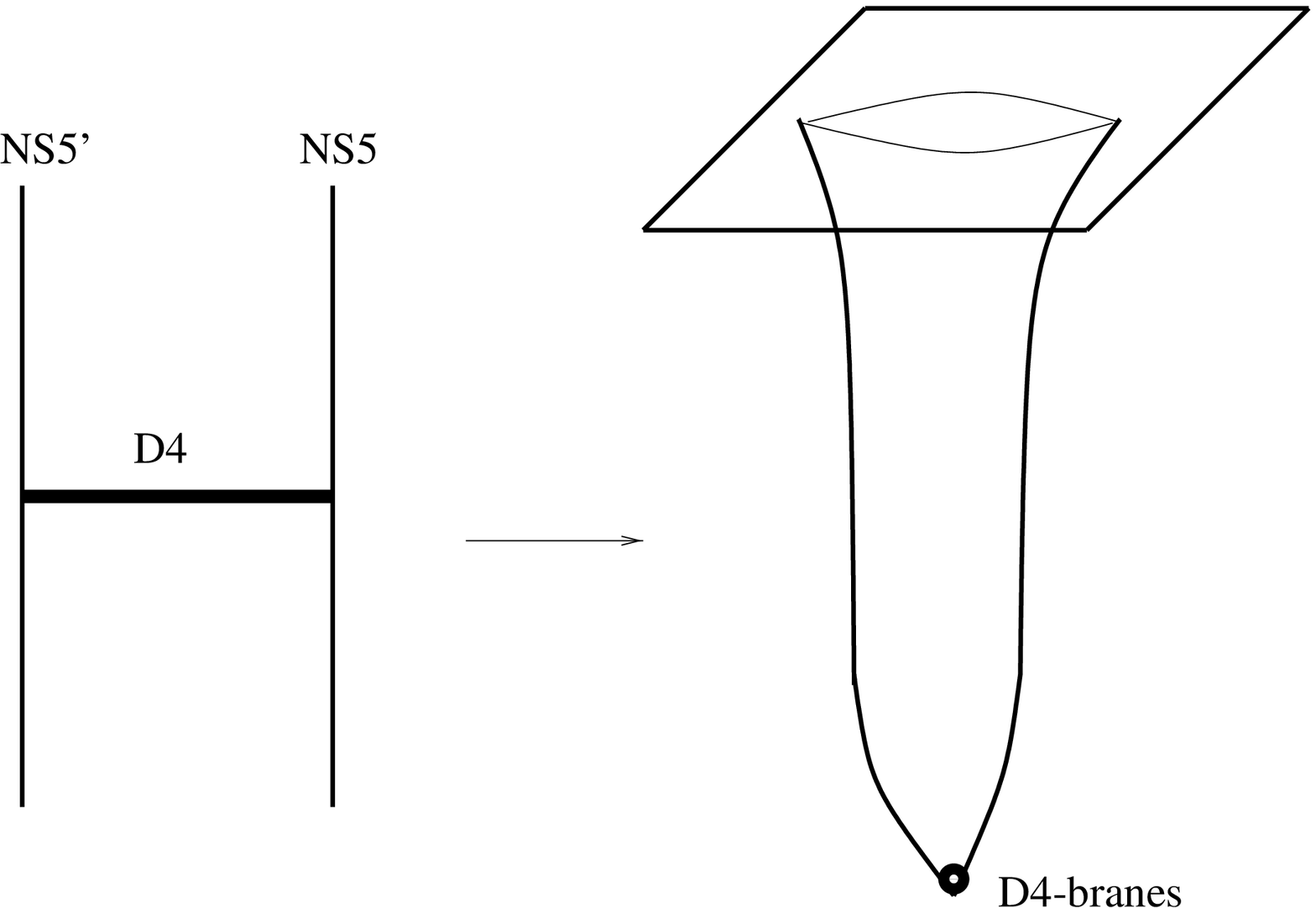}}
\vskip .1in
\noindent
Fig.\ 2: {\it The decoupled theory on D4-branes stretched between
NS5-branes is dual to D-branes near the tip of the cigar background.}
\vskip .5cm

Consider the brane configuration in figure 2 (for a review, see \gkrev).
An NS fivebrane is separated a distance $\ell$ from
a (possibly differently oriented) NS' fivebrane, and $N_c$ D4-branes
are stretched between them.
The low energy theory on the $R^{3,1}$ directions common to all the branes
is an $SU(N_c)$ gauge theory. The amount of supersymmetry
of the theory depends on the relative orientation of the
fivebranes. For instance, if the fivebranes are parallel
the four-dimensional theory is $N=2$ supersymmetric; a certain relative
rotation of the fivebranes breaks it to $N=1$ \egk.
The (classical) YM coupling is
\eqn\gym{g_{YM}^2={g_sl_s\over \ell}=g_{lst}(\ell)~.}
Following the discussion of the previous subsection, as $g_s\to 0$
(as well as $\ell/l_s\to 0$ such that $g_{YM}^2=g_{lst}$ is held fixed),
the decoupled theory is dual to a theory of
D-branes near the tip of a cigar geometry (see figure 2)
with $g_s(tip)\equiv g_{lst}(\ell)$ given in \gym.
At energies of the order $M_s$ the spectrum becomes similar to
that of LST: in the $N=2$ case it is like a six-dimensional LST on
$R^{5,1}\times SL(2)_2/U(1)$, while in the $N=1$ case
it is similar to a four-dimensional LST on
$R^{3,1}\times SL(2)_1/U(1)$.

We can add matter to the theory in several ways. One way is
to add D6-branes to the configuration and another is to add
D4-branes on the other sides of the NS5-branes. In the
second case the group that lives on the D4-branes in the
center is called the ``color group,'' whereas the
one on the D4-branes sticking to the right (and/or to the left)
is called the ``flavor group.''
If the number $N_c$ of colors equals
(up to a model dependent numerical factor)
the number of flavors $N_f$,
the four-dimensional theory is conformal
(in which case \gym\ is the exact gauge coupling).
Moreover, we compactify the space transverse to
$R^{3,1}$ on a six-dimensional space with
volume $V_6$.~\foot{Once this is done one should take care of global issues
by allowing the appropriate total number of NS5-branes,
introduce orientifolds, anti-D-branes, etc.
(see comments below); we assume that this is done.}
The four-dimensional Newton's constant is
\eqn\gn{{1\over G_4}\sim M_P^2 \sim {V_6\over g_s^2 l_s^8}~.}
\lref\egkrs{S. Elitzur, A. Giveon, D. Kutasov, E. Rabinovici and G.
Sarkissian, hep-th/0005052.}%
\lref\egkrsh{S. Elitzur, A. Giveon, D. Kutasov, E. Rabinovici and A.
Schwimmer,
Nucl.\ Phys.\ {\bf B505} (1997) 202, hep-th/9704104.}%

Consider turning one of the NS fivebranes in the
configuration above into a stack of fivebranes. Separating
these fivebranes a distance $L$ (say, on a circle
transverse to the D4-branes) corresponds in the low energy
four-dimensional SYM theory to changing certain parameters
in the superpotential \refs{\egk,\egkrsh}. Such
configurations allow to consider the physics of
color-flavor open strings which are bound to the stack of
NS5-branes perturbatively in $g_{lst}(L)=g_sl_s/L$ \gtip.
Such open strings in the background of NS5-branes were
studied in \egkrs. In particular, observables corresponding
to ``quarks'' ($(N_c,N_f)$ multiplets) and their excitations
were identified and their two point functions were computed
using the idea of holography, following \refs{\gk,\gktwo}.
The spectrum of charged particles in such four-dimensional
theories is thus very similar to the spectrum discussed in
the previous subsection: the massless particles correspond
to quarks, followed by a discrete Hagedorn spectrum with
masses of the order $M_s$, and a continuum. Following the
disk computations in \egkrs\ for the three point functions
one expects, in particular, that a massive color-color
string excitation can decay into a pair of
quark-anti-quark.

\lref\muk{S. Mukhi and N.V. Suryanarayana,
hep-th/0003219, JHEP {\bf 0006} (2000) 001.}
\vskip .1in
\noindent
Comments:
\item{1.}
The configurations discussed above are only part of a globaly consistent
brane configuration which includes more fivebranes, orientifolds,
D-branes and anti-D-branes. All these extra objects can be located a
distance of the order $l_s$ away from the configurations above, hence
physics due to the additional structure will show up at energies
above the string scale.
\item{2.}
Some systems of D4 and anti-D4-branes stretched between
non-parallel NS5-branes are expected to be stable non-BPS
brane configurations \muk. One might expect the
supersymmetry breaking scale $M_{SUSY}$ in such theories
to be of the order of the string scale:
$M_{SUSY}\sim M_s$.
\item{3.}
The system discussed in this subsection is related
to systems discussed in the previous subsection:
there is a U-duality relating D4-branes intersecting NS5-branes
to NS5-branes wrapped on a Riemann surface, which is
holographically dual to the string theory
on (generalizations of the) backgrounds of the form \four.

\newsec{Phenomenology of TeV LST}

\lref\KK{I. Antoniadis, K. Benakli and M. Quir\'os,
Phys. Lett. {\bf B331} (1994) 313; for a recent review see
I. Antoniadis and  K. Benakli, hep-ph/0007226 and references therein.}
\lref\indirect{E. Accomando, I. Antoniadis and  K. Benakli,
Nucl. Phys. {\bf B579} (2000) 3;
S. Cullen, M. Perelstein and M.E. Peskin, hep-ph/0001166;
D. Bourilkov,  hep-ph/0002172.}
\lref\mayr{P. Mayr, hep-th/9610162, Nucl. Phys. {\bf B494} (1997) 489.}

Here, we discuss the main phenomenological implications of the
above theories when the string tension and compactification scales
are in the TeV region. Because of the tiny value of the string
coupling $g_s\sim M_s/M_P\simeq 10^{-16}$, for all low energy
consequences we can take the limit $g_s\to 0$, in which case
gravity decouples and one is left over with LST having two
sectors. The non-abelian (Standard Model) particles confined on
NS5-branes (described by D-branes stretched between the NS5-branes)
or on D-branes stretched between NS5-branes (described by open strings
ending on the D-branes),
and a sector of gauge singlet closed little strings trapped in the
vicinity of the NS5-branes. Thus, there are three types of
possible excitations revealing new physics:
\item{(1)}
KK and winding modes of ordinary particles, signaling new
dimensions at a TeV.
\item{(2)}
String oscillations of the quarks, leptons and familiar gauge bosons.
\item{(3)}
Vibrational excitations of the little string at the TeV scale.
These are unique to this framework.

There are three types of theories,
each with different signatures: type IIB, type IIA
and theories with D-branes stretched between NS5-branes. We will consider
them in turn.
\hfill\break

Type IIB models:\hfill\break
In these the Standard Model gauge interactions are described by
a six-dimensional theory of
0-branes obtained as endpoints of D-strings on the NS5-branes. Their
tension $T_0$ is determined by eq. \mwb\ and can be identified with the
mass of the W-boson, $T_0\equiv T_{D1}L\simeq M_W$.~\foot{In theories with
large supersymmetry there are also magnetically charged particles, however,
those can be pushed above the $M_s\sim$ TeV scale.
More realistic models where supersymmetry is broken down to $N=1$
or $N=0$ can be obtained, say, by appropriate orbifoldings; in such theories
magnetically charged particles are projected out.}
The four-dimensional gauge coupling \gymf\ is determined by the area of
the two-dimensional compact space along the NS5-branes, which implies that
the compactification scale is an order of magnitude lower than the string
scale:
\eqn\compsc{M_c\equiv {1\over R_5}\sim {1\over R_4}=g_{YM}M_s\, .}
It follows that the first effects of charged particles beyond the
Standard Model that would be encountered in particle
accelerators are due to the production of KK excitations in the two extra
dimensions \KK. Neutral states will also appear at the TeV range; we
shall discuss them below.
\hfill\break

Type IIA models:\hfill\break
In these the gauge degrees of freedom are
described by strings, obtained as endlines
of D2-branes on the NS5-branes. Their tension $T_1\equiv T_{D2}L$ can
be obtained by T-duality from type IIB along, say, the direction $X^4$, so
that the gauge coupling \gymf\ is given by the ratio of the two radii.
As a result, $T_1=T_0/R_4\equiv M_W/R_4$ leading to
\eqn\stringt{T_1={M_W\over R_4}={M_W M_c\over g_{YM}^2}\, .}
On the other hand, $T_1=T_{D2}L=M_s^3L/g_s$, which combined with
eq.~\gtip\ yields:
\eqn\glstab{{\rm type \,\, IIA:}\qquad\qquad
g_{lst}={M_s^2\over T_1}~.}

{}From eq.~\stringt, $M_W$ is identified with the dual
compactification scale along the $X^4$ direction with
respect to the charged string tension $T_1$. If the
fundamental string tension $M_s^2$ is lighter than $T_1$,
then $M_s^2R_4<T_1R_4=M_W$ and closed little strings have
windings at energies lower than $M_W$, which is excluded
experimentally. Thus, the tension $T_1$ of the charged
gauge states is less than that of the little strings,
${\sqrt T_1}<M_s$, and in the energy interval between the
two tensions we will have an effective superconformal
theory of tensionless strings (see also \mayr). 
However, from eq.~\glstab,
$g_{lst} >1$ and we cannot reliably compute in little
string perturbation theory. It follows that in this case
the first effects of charged particles that would be
encountered in particle accelerators are KK modes of one
dimension along the $X^5$ direction and/or charged string
excitations with tension $T_1$, depending whether
$M_c\equiv R_5^{-1}$ is less or bigger than $M_W/g_{YM}^2$.
In both the type IIA and IIB frameworks the coupling of the
little strings to the standard model matter is unknown.
\hfill\break

D-branes in LST:\hfill\break
In type II models where
gauge interactions emerge from D-branes stretched between
NS5-branes, the low energy physics is described by the theories on
D-branes in LST, discussed in section 3.2. Note that the
mass of W bosons corresponds now to the separation of the D-branes
and is independent of separations of NS5-branes.

In this case, Standard Model particles have charged excitations
due to windings of open strings in the
directions transverse to the D4-branes
but along the two extra compact dimensions of the NS5-branes.
The energy of these excitations is
$M_s^2R_c$, where $R_c$ is the compactification scale.
If $R_c<l_s$ we can T-dualize $R_c\to\tilde R_c=l_s^2/R_c>l_s$.
In this case the D4-branes turn into D5-branes wraped on the compact
direction $\tilde R_c$, and $g_{lst}(\ell)$ in eq. \gym\ turns
into $\tilde g_{lst}(\ell)=g_{lst}(\ell)l_s/R_c$.
Charged excitations in this direction correspond now to KK modes
of open strings which are somewhat lighter than the string scale.
In fact, the weak coupling condition $\tilde g_{lst}<1$ gives
$l_s>R_c>g_{lst}l_s=g_{\rm YM}^2l_s$. Thus,
both the compactification and the string scales
are in the TeV region and in all these cases the energy of
such charged particles is around the TeV scale while little strings
are weakly coupled.

There are also KK modes of open strings in the direction along
which the D4-branes are stretched as well as windings along
the directions transverse to both the D4 and the NS5-branes;
those are very weakly coupled (and decouple in the $g_s\to 0$ limit).
In addition, there are of course fundamental open string
oscillator modes that are also charged under SM gauge interactions
and have TeV masses.
\hfill\break

The common thread of all three cases is the existence of a neutral sector
described by closed fundamental little strings that survive in the limit
$g_s\to 0$. They have non-trivial interactions among themselves with a
coupling $g_{lst}$ given in eq.~\gtip. Perturbative computations can
therefore be trusted when $g_{lst}<1$. Using eqs.~\glstab,
\stringt\ and \mwb, one obtains:
\eqn\glstabb{\eqalign{{\rm type \,\, IIA:}&\qquad\qquad\qquad
g_{lst}={M_s^2\over T_1}=g_{YM}^2{M_s^2\over M_W M_c}~,\cr
{\rm type \,\, IIB:}&\qquad\qquad\qquad g_{lst}={M_s\over M_W}~,\cr
{\rm D-branes\,\, in\,\, LST:}&\qquad\qquad\qquad
g_{lst}=g_{YM}^2={g_s\over \ell M_s}~.\cr}}
Recall that for theories of D-branes in the
presence of NS5-branes, $g_{lst}$ is
independent of the W boson mass which is determined by the
separation of the D-branes and not of the NS5-branes.

It follows that the discussion
of the perturbative spectrum of section 3 is strictly
speaking valid for the theories of D-branes in LST
if the separation of NS5-branes is larger than $g_sl_s\sim 1/M_P$.
The little string excitations can be produced in particle accelerators
if they dispose sufficient energy, or they can lead to
indirect effects in various processes, as the effects of
TeV string models based on type I theory \indirect.

The perturbative spectrum occurs at \refs{\gk,\gktwo}
\eqn\mnm{M_{n,m}^2={2M_s^2\over k}(n-1)(2m-n)~, \qquad
2m+1>2n>2m+1-k~, \qquad n\in Z~.}
The pole at $n=1$ corresponds to the light SM particle.
The other poles at $M_{n,m}^2\sim M_s^2/k$ are KK-type excitations,
due to the asymptotic radius of the cigar.
Each set of poles on the up-side-down parabola \mnm\ is followed by
a branch cut starting at the maximum of the parabola, that we discussed
in the previous section.

Similarly, observables corresponding to string excitations $N$
create from the vacuum particles with masses
\eqn\mex{M^2_{n,m;N}=M_{n,m}^2+NM_s^2~.}
Hence, each of the particles in \mnm\ is followed by a Regge
trajectory of string excitations~\foot{For charged open little strings
(open strings in the background of NS5-branes), some factors of 2 should
be added in eqs. \mnm, \mex\ relative to the neutral closed little strings
sector; see for instance eqs. (B.14), (B.15) in \egkrs.}.

\lref\ms{J.M. Maldacena and A. Strominger,
hep-th/9710014, JHEP {\bf 9712} (1997) 008.}%
\lref\bh{O. Aharony and T. Banks, hep-th/9812237,
JHEP {\bf 9903} (1999) 016.}%
\lref\ho{T. Harmark and N.A. Obers, hep-th/0005021,
Phys. Lett. {\bf B485} (2000) 285.}%
\lref\br{M. Berkooz and M. Rozali, hep-th/0005047,
JHEP {\bf 0005} (2000) 040.}%
\lref\kush{D. Kutasov and D.A. Sahakyan, hep-th/0012258.}%

It is interesting to consider the thermodynamics of LST at a TeV;
this can be done using its holographic description
\refs{\ms,\bh,\ho,\br,\kush}.
Strongly coupled LST has a Hagedorn density of states and the
Hagedorn temperature is $T_H=M_s/(2\pi\sqrt{k})$ \ms.
At high energy the entropy is
\eqn\enten{S=\beta_H E+\alpha \log E + O(1/E)~,}
leading to the temperature-energy relation
\eqn\temen{\beta={\partial S\over\partial E}
=\beta_H+\alpha/E+O(1/E^2)~.}
The sign of $\alpha$ indicates if $T_H=1/\beta_H$ is a limiting temperature
(where the energy density diverges as $T$ approaches $T_H$ from below)
or a temperature where a phase transition might occur.
Recently, $\alpha$ was computed and was shown to be negative \kush.
This suggests that a phase transition is expected at $T\sim M_s$,
similar to QCD.
The nature of the high temperature behavior of the theory might have
some interesting consequences in the physics of the early universe.

Weakly coupled LST ($g_{lst}<1$) has of course the same
high energy thermodynamics (when $E\gg M_s/g_{lst}$).
However, at intermediate energies $M_s<E<M_s/g_{lst}$
the weakly coupled little string excitations possess a
Hagedorn density of states with $T_H=M_s/(2\pi)$ \gk.

\lref\akt{I. Antoniadis, E. Kiritsis, T.N. Tomaras,
hep-ph/0004214.}%

Finally, we remark that gauge coupling Unification, the one
concrete quantitative success of the supersymmetric standard model
\dg, can be accommodated in a way parallel to \akt. There it was
shown that under general conditions, placing the color and weak
interactions on two different sets of branes (extended in different
directions in the internal compact space) implies one relation
among the three gauge couplings which naturally leads to the correct
value of the weak mixing angle, provided that we choose the fundamental
scale to be at a few TeV.

\lref\ds{N.V. Krasnikov, Phys. Lett. {\bf B193} (1987) 37;
M. Dine and Y. Shirman, Phys. Rev. {\bf D63} (2001) 046005,
hep-th/9906246, and references therein.}

\newsec{The Hierarchy Problem}

\lref\disei{M. Dine and N. Seiberg, Phys. Lett. {\bf B156} (1985) 55.}%

In the framework we described here, the hierarchy between the
Planck and the string scales is attributed to the smallness of the
string coupling.
A small string coupling is rather natural when supersymmetry is broken
due to the runaway potential generated for the dilaton \disei.
In a usual scenario where the YM couplings are of the order $g_s$
such a runaway behavior is a serious problem.
On the other hand, in the LST scenario considered in this note,
$g_{YM}$ is determined by geometrical data,
while $g_s$ is an independent parameter.
Yet, although $g_s\ll 1$ is required to set
the observed Newton's constant,
it should not runaway all the way to 0.
In this section, we describe a mechanism
determining the expectation value (VEV) of the dilaton and the
conditions for generating the desired hierarchy.

\lref\adm{N. Arkani-Hamed, S. Dimopoulos and J. March-Russell,
hep-th/9809024.}%
\lref\mayrtwo{P. Mayr, hep-th/0006204, JHEP {\bf 0011} (2000) 013.}%

Let us first remark that if the value of the string coupling is
chosen to be very small by hand, the resulting hierarchy is
obviously stable under radiative corrections even around a
non-supersymmetric string vacuum. This should be contrasted  with
the large dimension framework \add\  where stability of hierarchy
requires that massless bulk fields propagate in more than one large
compact dimensions \ab. Moreover, the vacuum energy in a
non-supersymmetric vacuum of the theories we described here behaves at most
as $M_s^4\sim ({\rm TeV})^4$ 
(see also \mayrtwo\ and the discussion below).
This should be again
contrasted with the framework of large dimensions which suffers in
general from the usual quadratic divergences $\sim M_s^2M_P^2$, unless
the bulk is supersymmetric \refs{\add,\adm}.

Dynamically determining the dilaton by minimizing an
effective potential faces the following problem. Since the
dilaton plays the role of string loop expansion parameter,
a generic non-trivial potential would mix several orders of
perturbation theory (as well as eventually non-perturbative
effects) and, in general, the minimum would be at a point
where different powers of $g_s$ compete and, as a result,
perturbation theory is unreliable. Moreover, the value of
the coupling is in general expected to be of order unity. A
possible exception using non-perturbative contributions
such as several condensates \ds\ appears very unnatural in
our case, since non-perturbative factors are extremely
suppressed in the desired very weak coupling limit.

One way to evade this problem is through the appearance of
logarithms. These can arise from loops of particles having
gauge interactions with masses depending on the string
coupling. The first difficulty is that gauge theories on
NS5-branes are independent of the string coupling. One
should therefore introduce a new gauge (hidden) sector
living on D-branes and thus having a gauge coupling given
by $g_s^{1/2}$. The second difficulty is that massive
particles on D-branes have in general masses set by their
separation, their motion or the string scale itself (for
string excitations) all of which are independent of $g_s$.
In these cases, loop effects cannot produce logarithms of
$g_s$. However when masses are induced radiatively, they
depend on the string coupling and can give rise to logs.
One such example arises when there is an anomalous $U(1)$.
The anomaly is cancelled by an appropriate shift of an
axion from the Ramond-Ramond sector and the abelian gauge
field acquires a mass $m_A=g_s^{1/2}M_s$. Integrating out
this field, one obtains a potential term proportional to
$m_A^4\ln m_A$, or equivalently (in the string frame):
\eqn\veff{V_{\rm eff}=g_s^2(v_1\ln g_s+v_2)M_s^4+cM_s^4\, ,}
with $v_{1,2}$ and $c$ numerical constants. The first two terms proportional
to $v_{1,2}$ correspond to two string loops contributions (genus 2), while
$c$ arises at the one loop (genus 1).

The effective potential \veff\ has an extremum at
\eqn\extr{\langle g_s\rangle=e^{-1/2-v_2/v_1}\, ,}
which is a minimum when $v_1$
is positive. This minimum can be exponentially small when $v_2$ is
just one or two orders of magnitude bigger than $v_1$, which is
not unreasonable since $v_1$ is determined entirely from the loop
of the anomalous $U(1)$ while $v_2$ receives contributions from
all string modes.

Note that in general, in the presence of D-branes, one may expect
an additional contribution to $V_{\rm eff}$ proportional to $g_s$,
arising from genus 3/2. Such a term would destabilize the minimum
\extr\ and is assumed to vanish. In fact this condition is related
to the problem of fine tuning the cosmological constant.  In the
above example \veff, we should therefore impose
\eqn\const{c=-\langle g_s\rangle^2(v_1\ln \langle g_s\rangle+v_2)
+{\cal  O}(\langle g_s\rangle^4)\, .}

Another example of logarithmic corrections to the potential may be
provided in models of the Coleman-Weinberg type, where a
classically massless scalar field with a tree-level potential
acquires a non-trivial VEV driven by a negative squared mass
generated radiatively. In this case, the scalar potential takes
the form $V_{\rm eff}\sim \Phi^4/g_s-\mu^2\Phi^2$, up to $\ln\Phi$
corrections in both terms. Then, its minimization fixes
$\langle \Phi\rangle\propto g_s^{1/2}$
and leads to $V_{\rm eff}\sim g_s\ln g_s$ which is
similar to the expression \veff\ that we studied above.

An alternative possibility of fixing the dilaton without
generating a potential would be during the cosmological evolution
of the universe, following the suggestion of Damour and Polyakov
\dap. The basic requirement is that all couplings and masses of
the effective theory should depend on the dilaton through
the same function. If in addition this function has an
extremum, the cosmological evolution will ``push" the
dilaton towards this extremum. This happens during matter
dominated era, crossing mass thresholds in radiation
dominated, as well as during any period of inflation. As a
result, the dilaton couples quadratically to matter and its
mass can vanish without causing any dangerous long range
force.

The main requirement of a universal functional dependence
seems however very unlikely to be satisfied in heterotic and type I
string vacua. On the contrary, theories on NS5-branes seem to provide
a natural framework for realizing such a requirement, since in the
string frame the matter action is independent of the dilaton while
graviton kinetic terms may acquire a non-trivial dilaton dependence:
\eqn\dampol{{\cal L}=F(g_s)R+{\cal L}_{\rm matter}\, ,}
where $F(g_s)=1/g_s^2+$ higher order and non-perturbative
corrections. It follows that upon rescaling the metric into
the Einstein frame, all mass parameters of the matter
Lagrangian will depend on the single ``universal'' function
$F$. It is not however clear under what conditions $F$
would have an extremum at a tiny value of $g_s$.

\lref\dn{T.~Damour and K.~Nordtvedt,
Phys.\ Rev. {\bf D48} (1993) 3436.}%
\lref\skw{D.~I.~Santiago, D.~Kalligas and R.~V.~Wagoner,
gr-qc/9706017, Phys.\ Rev. {\bf D56} (1997) 7627.}%

The dilaton may approach but not precisely reach the extremum of
$F$ in cosmological time \dn. This results in a small universal
linear coupling of the dilaton to matter -- but not to radiation --
proportional to the fractional deviation $\alpha$ of the dilaton's
present position away from its minimum. Such a scalar admixture to
gravity has several possible observational consequences, including
the bending of light and the Shapiro time delay of signals 
\refs{\dn,\skw}. The most stringent bounds come from primordial
nucleosynthesis, and they constrain the present value of $\alpha$
to be less than a few percent \skw. A precision test, possibly
improving the present limit of $\alpha$ by over an order of
magnitude, will take place in the relativistic gyroscope (or
Gravity Probe B) experiment that will be launched in 2002.

In general, small flavor-dependent effects are expected to spoil
the exact universality of the coupling of the dilaton through the
function $F$. In the string frame these are expected to show up as
small $g_s$-dependent corrections to the various gauge-invariant
terms in the matter Lagrangian of eq. \dampol.
This results in a linear coupling of the dilaton to matter which
is flavor-dependent and, therefore, leads to violations of
the principle of equivalence -- estimated to be proportional to
the product of $\alpha \times g_s$. These are potentially observable
in the upcoming satellite experiment STEP, which will test the
principle of equivalence to one part in $10^{18}$ \dap. Since
$g_s$ grows with $M_s$ and $V_6$, the predicted violations are
larger for string scale above a TeV or bulk volume above a TeV$^{-6}$
(see footnote 2).

\newsec{Remarks}

We end with some comments, first on the cosmological constant. In
models with infinitesimal string coupling, it seems
that the vacuum energy may be consistent with the present
experimental bound if the perturbative contributions are
arranged to vanish in one and two loops, while
non-perturbative corrections appear to be extremely
suppressed. Indeed, the three loop contribution is of order
$g_s^4M_s^4\sim M_s^8/M_P^4$ which is just of the order
suggested by present observations for $M_s\sim 1$ TeV.
This may provide a new framework for explaining the
smallness of the cosmological constant which deserves
further investigation.

The theory of NS fivebranes with the string scale set equal
to the electroweak scale and with a very small asymptotic
string coupling realizes several recent ideas in explicit
string theory backgrounds.

{}For example, the tip of the cigar is a concrete realization
in string theory of what one would call in \rs\
``the $d$-dimensional negative tension brane'' (see figure 1).
Unlike possible
realizations of warped compactification scenaria in string
theory, here the theory at high energies is {\it not} a CFT;
it is a string theory with a scale $M_s$ coupled (weakly) to
gravity. Nevertheless, there is an analogy between the
dilaton and the $y$ coordinate responsible for the
exponential hierarchy in warped compactification scenaria.

\bigskip
\noindent{\bf Acknowledgements:}
We would like to thank Nima Arkani-Hamed, Micha Berkooz, 
Thibault Damour, Nemanja Kaloper,
David Kutasov, Peter Mayr, Sasha Polyakov, Massimo Porrati,
Eva Silverstein, Herman Verlinde, Robert Wagoner and Edward Witten 
for valuable discussions. AG thanks the Theory Division at
CERN, the Ecole Polytechnique and the ITP at Stanford
University for their warm hospitality. This work is
supported in part by the EEC TMR contract ERBFMRX-CT96-0090
and RTN contract HPRN-CT-2000-00122, and in part by the
INTAS project 991590. The work of SD is supported by NSF
grant PHY-9870115-003. The work of AG is supported in part
by the BSF -- American-Israel Bi-National Science
Foundation, by the GIF
-- German-Israel Foundation for Scientific Research,
and by the Israel Academy of Sciences and Humanities -- Centers of
Excellence Program.

\listrefs
\end